\documentclass[11pt,a4paper]{article}
\usepackage{amssymb} 
\usepackage{feynmf}

\newcommand{\Tr}{\mathop{\rm Tr}\nolimits}

\def \no {\nonumber}
\def \be {\begin{equation}}
\def \ee {\end{equation}}
\def \bea {\begin{eqnarray}}
\def \eea {\end{eqnarray}}

\def \nc {noncommutative }

\begin{document}
\begin{fmffile}{fmfncren}

\baselineskip 0.55 cm
\begin{flushright}
SISSA 105/00/EP\\
hep-th/0011088
\end{flushright}
\begin{center}

{\Large{\bf Renormalization of \nc $U(N)$ gauge theories }}
\vskip .5cm

{\Large  L.Bonora, M.Salizzoni}

\vskip .5cm
{bonora@sissa.it, sali@sissa.it}
\vskip .5cm

{\it Scuola Internazionale Superiore di Studi Avanzati,

Via Beirut 2-4, 34014 Trieste, Italy, and\\

INFN, Sezione di Trieste}

\end{center}

\vskip 2cm 
\begin{abstract}
We give an explicit proof that \nc $U(N)$ gauge theories are one--loop
renormalizable.
\end{abstract}
\newpage   
\section{Introduction}

Since \cite{Bound} it has been clear that the concept of space in the presence
of nearby D--branes is radically modified. One way to express it is to say that
the positions of the branes are replaced by suitable matrices, whose entries
actually represent open strings stretched among them. This entails in particular
that at short brane distances space becomes \nc. This picture is indeed
very suggestive, but not very effective in representing space 
non--commutativity. It has been only recently, \cite{HD,AAS,Chu,Dir,SW}, that noncommutativity
has surfaced in a very effective and manageable way. This happens precisely
when D--branes are in presence of a constant NSNS B-field. In this case,
the low energy effective action of the open strings attached to the branes
can be represented by a Euclidean field theory defined on a \nc spacetime. 
All this clearly holds at a semiclassical level (i.e. tree amplitudes
computed in the string and the field theory setting compare well). 
However one can
try to compare loop amplitudes calculated both in string theory and in the
corresponding \nc field theory, in order to see how effective the \nc
effective field theory is. Several calculations of this type have been carried
out, see \cite{filk,VGB,CDP,MSR,ABK,tomas,CHMS,haw,GKMRS,BCR,CRS,CR,GGRS,GS,micu}
and in particular \cite{jabbari,tomas,adi}, and some 
discrepancies have recently surfaced, \cite{trek,zanon}. The latter
papers seem to imply that more general * products are 
necessary, in order for the effective field theory to faithfully represent
string loop contributions.

It seems to be important therefore to know exactly what are the properties
of a \nc YM theory we can rely on. One of the basic properties is  
renormalizability. In this paper we consider  
a \nc YM theory with $U(N)$ gauge group in 4D without matter, and study its 
one--loop renormalizability properties. Since we take for granted that 
non--planar singularities are dumped by the noncommutative parameter $\theta$ 
\cite{filk,MRS,haya,MST}, we only 
consider the planar one--loop contributions. This has already been partially
done as far as the $U(1)$ gauge theory is concerned \cite{jabbari,tomas},
and for two-- and three--point functions for $U(N)$, \cite{adi}. We want to 
complete such calculations by computing the one--loop four--point
function in a pure \nc $U(N)$ Yang--Mills theory in 4D, for which some doubts
have been raised. It is obvious that if gauge (BRST) invariance is assumed, the 
calculation we do is superfluous. But in a \nc gauge theory some caution has 
to be used and BRST invariance 
has to be proved rather then assumed. We have in fact a simple counterexample: 
NCSO(2) gauge theory, which was introduced in \cite{BSST}. If we define the
Feynman rules in the most obvious way, it is immediate to see that for 2- and
3-gluon vertices the one--loop corrections identically vanish, while
the 4-gluon has a divergent part at one-loop. Therefore this theory is 
not one--loop renormalizable (at least in the ordinary sense).

Our final result is however reassuring. {\it Noncommutative $U(N)$ gauge 
theories are one--loop renormalizable}.
           
The paper is organized as follows. In the next section we set notations
and conventions. In section 3 we compute the relevant one--loop amplitudes for
a \nc $U(N)$ theory. We also make a comment on the possibility of
defining consistent \nc $SU(N)$ Feynman rules.

\section{Notations, conventions and $u(N)$ tensors}

Our \nc theory is specified by the action
\be
S = \int d^4x \,\Tr \left(- \frac {1}{4}F_{\mu\nu}F^{\mu\nu}-\frac{1}{2\alpha}(\partial_{\mu} A^{\mu})^2 +\frac{1}{2}(i\bar{c}*\partial_{\mu}D^{\mu}c\,-\,i\partial_{\mu}D^{\mu}c*\bar{c})\right) \label{action}
\ee
where
\be
F_{\mu\nu}= \partial_\mu A_\nu- \partial_\nu A_\mu -ig(A_\mu * A_\nu
 -A_\nu*A_\mu)\label{Fmunu}
\ee
and the Moyal product is defined 
with respect to the parameter $\theta^{\mu\nu}$.
The potential $A_\mu$ is valued in the Lie algebra $u(N)$, i.e. is an 
hermitian matrix, and we will choose the Feynman gauge $\alpha=1$. 
As is customary in dealing with 4D field theories, 
throughout the paper we use a Minkowski formulation of the theory,
although its brane origin is Euclidean. 

Since the properties of the Lie algebra $u(N)$ tensors are crucial in our
calculation, we devote the rest of this section to deriving them.
 
We use a basis $t^a$, $a=1,\ldots, N^2-1$ of traceless hermitean matrices 
for the Lie algebra $su(N)$, with normalization
\be 
{\rm tr} (t^a t^b) = \frac 12\delta^{ab}
\ee  
and structure constants $f_{abc}$ defined by
\be
[t^a, t^b]= i f_{abc}t^{c}~.
\ee
We define also the third order ad-invariant completely symmetric tensor 
$d_{abc}$ by means of
\be
\{t^a,t^b\}= \frac 1N \delta_{ab}+ d_{abc} t^{c}~.
\ee

Next we pass to the Lie algebra $u(N)$ by introducing the 
additional generator
$t^0= \frac{1}{\sqrt{2N}} {\bf 1}_N$. Corresponding to any index $a$ for $su(N)$
we introduce the index $A=(0,a)$, so that $A$ runs from $0$ to $N^2-1$.
We have
\be
[t^A,t^B]= i f_{ABC}t^C, \quad \{t^A,t^B\}= d_{ABC}t^C\label{fd}
\ee
where $f_{ABC}$ is completely antisymmetric, $f_{abc}$ is the same as for 
$su(N)$ and $f_{0BC}=0$, while $d_{ABC}$ is completely symmetric; 
$d_{abc}$ is the same as for $su(N)$,
$d_{0BC}=\sqrt{\frac {2}{N}} \delta_{BC}$, $d_{00c}=0$ and $d_{000}=
\sqrt{\frac {2}{N}}$. We have also
\be
{\rm Tr} (t^A t^B)= \frac 12 \delta^{AB}~.
\ee
The following identities hold and will be extensively used below
\bea
&&f_{ABX}\,f_{XCD}+ f_{ACX}\,f_{XDB}+f_{ADX}\,f_{XBC}=0\no\\
&&f_{ABX}\,d_{XCD}+ f_{ACX}\,d_{XDB}+f_{ADX}\,d_{XBC}=0\no\\
&&f_{ADX}\,f_{XBC}= d_{ABX}\,d_{XCD}- d_{ACX}\,d_{XDB}\label{fdiden}
\eea
Next we define the matrices $F_A, D_A$ as follows
\be
(F_A)_{BC} =f_{BAC},\quad\quad (D_A)_{BC}= d_{BAC} \label{FD}
\ee
In the evaluation of Feynman diagrams we need to know traces of two,
three and four such matrices. We borrow from the literature,
\cite{KR,Mac,Azc}, the corresponding results for $su(N)$ and extend them to
$u(N)$. Denoting by ${\widehat\Tr}$ the traces over the relevant $N^2\times N^2$
space, we obtain
\bea
&&{\widehat\Tr} (F_AF_B) = - N\,c_A\,\delta_{AB},\quad\quad c_A=1-\delta_{A,0}\no\\
&&{\widehat\Tr} (D_AD_B)= Nd_A\delta_{AB},\quad\quad d_A= 2-c_A\label{2tr}\\
&&{\widehat\Tr} (F_AD_B)=0 \no \\
&& \no\\
&&{\widehat\Tr}(F_AF_BF_C)= -\frac N2 \,f_{ABC}\no\\
&&{\widehat\Tr}(F_AF_BD_C)= -\frac N2 \,d_{ABC} \,c_A\,c_B\,d_C\no\\
&&{\widehat\Tr}(F_AD_BD_C)= \frac N2 \,f_{ABC}\label{3tr}\\
&&{\widehat\Tr}(D_AD_BD_C)= \frac N2\, \eta_{ABC}\,d_{ABC}\no
\eea
where $\eta_{ABC}= d_A\,d_B\,d_C - 4 \delta_{A+B+C,0}$. Finally
\bea
{{\widehat\Tr}}(F_AF_BF_CF_D)&=&\Big[ \frac 12 \delta_{(AB}
\delta_{CD)}+ \frac N8 \left(d_{ABX}\,d_{CDX}+ d_{ADX}\,d_{BCX}\right)\no\\
&&+ \frac N8 \left(f_{ADX}\,f_{BCX}-f_{ABX}\,f_{CDX}\right)\Big] 
c_A\,c_B\,c_C\,c_D\no\\
{\widehat\Tr}(F_AF_BF_CD_D)&=& - \frac N4 \left(d_{ABX}\,f_{CDX}+
f_{ABX}\,d_{CDX}\right)\,c_A\,c_B\,c_C\,d_D \no\\
{\widehat\Tr}(F_AF_BD_CD_D)&=&c_A\,c_B\Big[c_C\,c_D\frac 12 \left(\delta_{AC}\delta_{BD}-
\delta_{AB}\delta_{CD}+\delta_{AD}\delta_{BC}\right)\label{4tr}\\
&&+ \frac N8\, d_C\,d_D \left(f_{ABX}\,f_{CDX}- f_{ADX}\,f_{BCX}-
d_{ABX}\,d_{CDX}-  d_{ADX}\,d_{BCX}\right)\Big]\no\\
{\widehat\Tr}(F_AD_BF_CD_D)&=&\frac 12\left(\delta_{AB}\delta_{CD}+\delta_{AD}\delta_{BC}
-\delta_{AC}\delta_{BD}
\right)\,c_A\,c_B\,c_C\,c_D \no\\
&&+\frac N8 \left(f_{ABX}\,f_{CDX}- f_{ADX}\,f_{BCX}-d_{ABX}d_{CDX}-
 d_{ADX}\,d_{BCX}\right)\,c_A\,d_B\,c_C\,d_D \no\\
 {\widehat\Tr}(F_AD_BD_CD_D)&=& \frac N4 \left( f_{ABX}\,d_{CDX}+
d_{ABX}\,f_{CDX}\right)c_A\,d_B\,d_C\,d_D \no\\
 {\widehat\Tr}(D_AD_BD_CD_D)&=&\frac 12 \delta_{(AB}
\delta_{CD)}\,c_A\,c_B\,c_C\,c_D \no\\
&&+\frac N8 \left(f_{ADX}\,f_{BCX}-f_{ABX}\,f_{CDX}+d_{ABX}\,d_{CDX}+
 d_{ADX}\,d_{BCX}\right)\eta_{ABCD}\no
\eea
where $\eta_{ABCD}= d_A\,d_B\,d_C\,d_D -8 \delta_{A+B+C+D,0}$.

\section{Two--, three-- and four--point functions at one loop}

The Feynman rules are collected in Appendix. Evaluating the one--loop
contributions is lengthy but straightforward. In this section we consider
the planar part of the 2--, 3-- and 4--point functions and, adopting
the dimensional regularization ($\epsilon=4-D$, as usual), we extract first
the planar part and, out of it, the divergent part. 
The relevant results are written down below.
The 2-- and 3--point functions are exactly parallel
to the corresponding ones in ordinary gauge theories, and some of them are 
written down below only for the sake of comparison. 

Gluons carry Lorentz indices $\mu,\nu,...\,$, color indices $A,B,...\,$, and
momenta $p,q,...\,$. Ghosts carry only the last two type of labels. All the 
momenta are entering, unless otherwise specified, and we use the notation 
$p\times q = \frac{1}{2}p_{\mu}\theta^{\mu\nu}q_{\nu}$.

{\bf 2--point function}. We have two nonvanishing contribution to the
UV divergent part:

\noindent
-- gluons circulating inside the loop:
\begin{equation}
  i\frac{1}{(4\pi)^2} \frac{2}{\epsilon} \delta_{AB} N \left[ \frac{19}{12}
  g_{\mu\rho} p^{2}
  -\frac{11}{6} p_{\mu} p_{\nu}
  \right]
\end{equation}
-- ghosts circulating inside the loop:
\begin{equation}
  i\frac{1}{(4\pi)^2} \frac{2}{\epsilon} \delta_{AB} N
  \left[ \frac{1}{12} g_{\mu\rho} p^{2}
  +\frac{1}{6}p_{\mu}p_{\nu}
  \right]
\end{equation}
Their sum is:

\begin{equation}
  i\frac{1}{(4\pi)^2}\frac{2}{\epsilon}\delta_{AB}N \frac{5}{3}\left[g_{\mu\rho}p^{2}
  -p_{\mu}p_{\nu}
  \right]
\end{equation}
\noindent
which entails the usual renormalization constant 
\begin{equation}
Z_3=1 + \frac 53
g^{2}N\frac{1}{(4\pi)^{2}}\frac{2}{\epsilon}~~~.
\end{equation}

{\bf 3--point function}. The external gluons carry labels $(A,p,\mu)$, 
$(B,q,\nu)$
and $(C,k,\lambda)$ for the Lie algebra, momentum and Lorentz indices. They are 
ordered in anticlockwise sense. 
The triangle diagram gives
\begin{eqnarray}
-\frac{13}{8}
      g^{3}N\frac{1}{(4\pi)^{2}}\frac{2}{\epsilon}
      \left( \cos(p\times q) f_{ABC} + \sin (p\times q) d_{ABC} \right)
      \nonumber   \\
     \cdot \left((p-q)_{\lambda}g_{\mu\nu}+(q-k)_{\mu}g_{\nu\lambda}
      +(k-p)_{\nu}g_{\mu\lambda}
      \right)
      \label{tri}
\end{eqnarray}
The diagram with one three gluon vertex and one four--gluon vertex gives:
\begin{eqnarray}
    \frac{9}{4}g^{3}N \frac{1}{(4\pi)^{2}}\frac{2}{\epsilon}
      \left( \cos(p\times q) f_{ABC} + \sin (p\times q) d_{ABC} \right)
      \nonumber   \\
      \cdot \left((p-q)_{\lambda}g_{\mu\nu}+(q-k)_{\mu}g_{\nu\lambda}
      +(k-p)_{\nu}g_{\mu\lambda}
      \right)
\end{eqnarray}
The contribution of the two ghost circulating diagrams is:
\begin{eqnarray}
\frac{1}{24}
      g^{3}N \frac{1}{(4\pi)^{2}}\frac{2}{\epsilon}
      \left( \cos(p\times q) f_{ABC} + \sin (p\times q) d_{ABC} \right)
      \nonumber   \\
     \cdot \left((p-q)_{\lambda}g_{\mu\nu}+(q-k)_{\mu}g_{\nu\lambda}
      +(k-p)_{\nu}g_{\mu\lambda}
      \right)
      \label{ghost}
\end{eqnarray}
The sum of the coefficients is
\begin{equation}
      - \frac{13}{8} + \frac{9}{4} + \frac{1}{24} = \frac{2}{3}
\end{equation}
Therefore, as in the ordinary YM theory, the renormalization constant $Z_1$ 
is
\begin{equation}
Z_{1}=1+ \frac{2}{3}g^{2}N\frac{1}{(4\pi)^{2}}\frac{2}{\epsilon}~~~.
\end{equation}

\begin{figure}
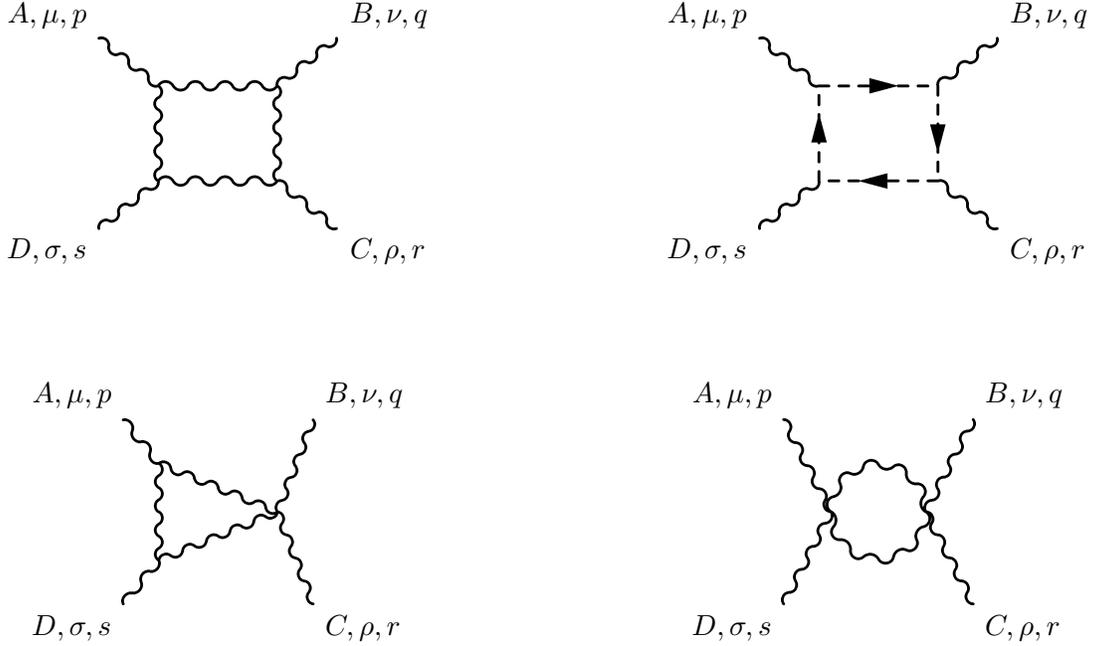

\begin{eqnarray*}
 ~~~~~~~~~~~~ & ~~\parbox{90mm}{
\begin{fmfchar*}(90,90)
   \fmftop{i1,i2}
   \fmfbottom{o1,o2}
   \fmf{photon}{i1,v1}
   \fmf{photon}{v4,i2}
   \fmf{photon}{o1,v2}
   \fmf{photon}{v3,o2}
   \fmf{photon,tension=.5}{v1,v4}
   \fmf{photon,tension=.5}{v2,v3}
   \fmf{photon,tension=.5}{v1,v2}
   \fmf{photon,tension=.5}{v3,v4}
   \fmflabel{$A, \mu, p$}{i1}
   \fmflabel{$C, \rho, r$}{o2}
   \fmflabel{$B, \nu, q$}{i2}
   \fmflabel{$D, \sigma, s$}{o1}
\end{fmfchar*}}
&\! \!\!\!\!\!\!\!\!\parbox{90mm}{
\begin{fmfchar*}(90,90)
   \fmftop{i1,i2}
   \fmfbottom{o1,o2}
   \fmf{photon}{i1,v1}
   \fmf{photon}{v4,i2}
   \fmf{photon}{o1,v2}
   \fmf{photon}{v3,o2}
   \fmf{scalar,tension=.5}{v2,v1}
   \fmf{scalar,tension=.5}{v1,v4}
   \fmf{scalar,tension=.5}{v4,v3}
   \fmf{scalar,tension=.5}{v3,v2}
   \fmflabel{$A, \mu, p$}{i1}
   \fmflabel{$C, \rho, r$}{o2}
   \fmflabel{$B, \nu, q$}{i2}
   \fmflabel{$D, \sigma, s$}{o1}
\end{fmfchar*}} 
\end{eqnarray*}
\bigskip

\begin{eqnarray*}
~~~~~~~~~~~~&~~\parbox{90mm}{
\begin{fmfchar*}(90,70) 
    \fmfleft{i1,i2}
    \fmfright{o1,o2}
    \fmf{photon}{i1,v2}
    \fmf{photon,tension=.3}{v3,v1}   
    \fmf{photon,tension=.3}{v1,v2}
    \fmf{photon,tension=.3}{v3,v2}
    \fmf{photon}{i2,v1}
    \fmf{photon}{o1,v3,o2}
    \fmflabel{$A, \mu, p$}{i2}
    \fmflabel{$C, \rho, r$}{o1}
    \fmflabel{$B, \nu, q$}{o2}
    \fmflabel{$D, \sigma, s$}{i1}
\end{fmfchar*}}
&
\!\!\!\!\!\!\!\!\!\parbox{90mm}{
\begin{fmfchar*}(90,70) 
    \fmfleft{i1,i2}
    \fmfright{o1,o2}
    \fmf{photon}{i1,v1,i2}
    \fmf{photon}{o2,v2,o1}
    \fmf{photon, right, tension=1/2}{v1,v2}
    \fmf{photon, left, tension=1/2}{v1,v2}
    \fmflabel{$A, \mu, p$}{i2}
    \fmflabel{$C, \rho, r$}{o1}
    \fmflabel{$B, \nu, q$}{o2}
    \fmflabel{$D, \sigma, s$}{i1}
\end{fmfchar*}}
\end{eqnarray*}
\vspace{10mm}
\caption{One loop contributions to the four-point function}
\end{figure}
\bigskip
\bigskip
{\bf 4--point function}. The external gluons carry labels $(A,\mu,p)$,
$(B,\nu,q)$, $(C,\rho,r)$ and $(D,\sigma,s)$ for Lie algebra, Lorentz index and
momentum, as shown in Figure~1.

There are four distinct graphs contributing to the 4--gluon vertex: the gluon
box $\mathfrak b$, the ghost box $\mathfrak g$, the gluon triangle $\mathfrak t$
and the gluon candy $\mathfrak c$. There are two main type of contributions,
distinguished by their Lie algebra tensor structure. The first is characterized by
Kronecker delta functions in the Lie algebra indices, while the second 
consists of $d$ and $f$ tensors. The first type contributions, which
are potentially dangerous for renormalizability, fortunately vanish.
 
The second type contributions have the general form
\bea
\!\!\!\!\!-ig^4 \frac{2}{\epsilon}\, \frac{1}{(4\pi)^2}&\!\!\Big[&
\!\!\!\Big(\frac N8 \,\cos(p\times s -q\times r)\, L_{ABCD} + \frac N8 \,
\sin(p\times s-q\times r)\, M_{ABCD}\Big)\,K^{\mathfrak i}_{\mu\nu\rho\sigma} \no\\
&\!\!+&\!\!\!\Big(\frac N8 \,\cos(p\times r -q\times s)\, L_{BACD} -   
\frac N8 \,\sin(p\times r -q\times s)\, M_{BACD}\Big)\,K^{\mathfrak i}_{\nu\mu\rho\sigma} \no\\
&\!\!+&\!\!\!\Big(\frac N8 \,\cos(p\times s +q\times r) \,L_{ACBD} + 
\frac N8 \,\sin(p\times s+
q\times r)\, M_{ACBD}\Big)\,K^{\mathfrak i}_{\mu\rho\nu\sigma} \Big]
\label{second}
\eea
where
\bea
L_{ABCD}&=& d_{ABX}\,d_{CDX}+d_{ADX}\,d_{CBX}-f_{ABX}\,f_{CDX}
+f_{ADX}f_{BCX}\no\\
M_{ABCD}&=&d_{ABX}\,f_{CDX}+d_{ADX}\,f_{BCX}+f_{ABX}\,d_{CDX}
-f_{ADX}\,d_{BCX}\label{LM}
\eea
and
 \bea
K^{\mathfrak b}_{\mu\nu\rho\sigma}&=& \frac{94}{3}\, g_{\mu\nu}\,g_{\rho\sigma}
+ \frac{94}{3}\, g_{\mu\sigma}\,g_{\nu\rho}+
\frac{34}{3}\, g_{\mu\rho}\,g_{\nu\sigma}
\no\\
K^{\mathfrak g}_{\mu\nu\rho\sigma}&=& -\frac{1}{3}\, (g_{\mu\nu}\,g_{\rho\sigma}
+g_{\mu\sigma}\,g_{\nu\rho}+g_{\mu\rho}\,g_{\nu\sigma})\label{Ki}\\
K^{\mathfrak t}_{\mu\nu\rho\sigma}&=& -(46\, g_{\mu\nu}\,g_{\rho\sigma}
+ 46 \,g_{\mu\sigma}\,g_{\nu\rho}-32\,g_{\mu\rho}\,g_{\nu\sigma})\no\\
K^{\mathfrak c}_{\mu\nu\rho\sigma}&=& 16 \,(7\,g_{\mu\nu}\,g_{\rho\sigma}
+7 \,g_{\mu\sigma}\,g_{\nu\rho}- 8\,g_{\mu\rho}\,g_{\nu\sigma})\no
\eea

The 4--point vertex is now easily calculated by summing all the contributions
with the appropriate symmetry factors. The contributions (\ref{second})  
give rise to 
\bea
\!\!\!\!
i\frac {N}{12} g^4 \frac{2}{\epsilon} \,\frac{1}{(4\pi)^2}&\!\!\Big[&
\!\!\Big( \cos(p\times s -q\times r)\, L_{ABCD} +  \sin(p\times s-
q\times r)\, M_{ABCD}\Big)\,T_{\mu\nu\rho\sigma} \no\\
&\!\!+&\!\!\Big( \cos(p\times r -q\times s) \,L_{BACD} -   
 \sin(p\times r -q\times s)\, M_{BACD}\Big)\,T_{\nu\mu\rho\sigma} \no\\
&\!\!+&\!\!\Big( \cos(p\times s +q\times r)\, L_{ACBD} + 
 \sin(p\times s+
q\times r)\, M_{ACBD}\Big)\,T_{\mu\rho\nu\sigma} \Big]\label{secondf}
\eea
where
\be
T_{\mu\nu\rho\sigma}= g_{\mu\nu}\,g_{\rho\sigma}
+\,g_{\mu\sigma}\,g_{\nu\rho}- 2\,g_{\mu\rho}\,g_{\nu\sigma}\label{T}
\ee

Comparing (\ref{secondf}) with eq.(\ref{4gluv}) in the Appendix, we see that
the contribution (\ref{secondf}) implies that the four--A term
in the action is renormalized with a $Z_4$ given by
\bea
Z_4 = 1 -\frac{1}{3}\, g^2\, N\, 
\frac{1}{(4\pi)^2}\frac{2}{\epsilon}~~~.\label{Z4}
\eea
This is the same renormalization that occurs in ordinary 
$U(N)$ Yang--Mills theories. Therefore, {\it the \nc $U(N)$ Yang--Mills theories 
are one--loop renormalizable}.  
 
The $U(1)$ case must be treated separately. Using the corresponding Feynman
rules (see Appendix), one finds the 2-- and 3--point contributions evaluated 
above with $f=0$ and $d=1$ and multiplied by $\frac 12$. As for the 4--point 
function, the term corresponding
to (\ref{secondf}) is obtained by setting $L=2$ and $M=0$ in the latter.
Therefore all the renormalization constants satisfy the renormalization 
conditions, and, as a consequence, the \nc $U(1)$ gauge theory is one--loop 
renormalizable too, \cite{jabbari,tomas}.

We would like finally to present some results (which are obtained without much
effort as byproducts of the previous calculations) concerning a restriction
from the $U(N)$ to the $SU(N)$ case. It is not known what a \nc $SU(N)$ gauge 
theory is, although an attempt of defining it has been done recently, 
\cite{MSSW}. In particular we do not know the explicit form of the action.
Therefore we can only try to guess the relevant Feynman rules. The most obvious
possibility one can envisage is that they are simply obtained from the Feynman 
rules of the \nc $U(N)$ theory by restricting everywhere the $U(N)$ indices 
$A,B,...$ to the corresponding $SU(N)$ ones $a,b,...$. As one can see  
in this case the renormalization constants do not coincide with 
the ones in the ordinary $SU(N)$ gauge theory. Strictly speaking this is not enough
to conclude that the \nc $SU(N)$ theory is nonrenormalizable, unless one assumes
that the $\theta\to 0$ limit of the quantum theory is smooth.

However, even allowing for such more general possibility, it is easy to show 
that the theory defined by such Feynman rules is not one--loop renormalizable, 
see also \cite{adi}.
To this purpose it is sufficient 
to compare the ratio of the  renormalization constants of gluon propagator, 
$Z_3$,  and the three gluon vertex, $Z_1$, with the ratio of ghost 
propagator, $\tilde{Z_3}$, and ghost-ghost-gluon vertex, $\tilde{Z_1}$. If 
the $SU(N)$ theory were renormalizable, we should find 
$Z_3/Z_1 = \tilde{Z_3}/\tilde{Z_1}$. Instead we obtain 
\begin{eqnarray*}
Z_1 & = & 1 + g^2 \frac{1}{(4\pi)^2}\frac{2}{\epsilon}\,\,
              \frac{1}{4} \left(\frac{N^2 - 2}{N}\right) \\
Z_3 & = & 1 + g^2 \frac{1}{(4\pi)^2}\frac{2}{\epsilon}\,\,
              \frac{5}{3} \left(\frac{N^2 - 2}{N}\right) \\
\tilde{Z_1} & = & 1 - g^2 \frac{1}{(4\pi)^2}
                \frac{2}{\epsilon}\,\,\frac{1}{2} \left(\frac{N^2 - 3}{N}\right) 
                \\
\tilde{Z_3} & = & 1 - g^2 \frac{1}{(4\pi)^2}
                \frac{2}{\epsilon}\,\,\frac{1}{2} \left(\frac{N^2 - 2}{N}\right)
                ~~~, \\
\end{eqnarray*}
where we used the traces over the $SU(N)$ indices that can be found in 
\cite{Mac}.
 
\section*{Appendix. Feynman rules for \nc $U(N)$ theories.}

Gluons carry Lorentz indices $\mu,\nu,...$, color indices $A,B,...$, and
momenta $p,q,...$. Ghosts carry only the last two type of labels. All the 
momenta are entering unless otherwise specified.
\vskip .2cm
\noindent{\bf gluon propagator}. 
\begin{eqnarray}
~~~~\parbox{30mm}{
\begin{fmfchar*}(60,30)
   \fmfleft{i}
   \fmfright{o}
   \fmf{photon, label=$p$, labe.side=left}{i,o}
   \fmflabel{$A,\mu$}{i}
   \fmflabel{$B,\nu$}{o}
\end{fmfchar*}}~~~&~~~&~~~ -\frac {i}{p^2} \delta_{AB} g_{\mu\nu}
\end{eqnarray}
\vskip .2cm
\noindent{\bf ghost propagator}.
\begin{eqnarray}
\parbox{30mm}{
\begin{fmfchar*}(60,30)
   \fmfleft{i}
   \fmfright{o}
   \fmf{scalar, label=$p$, labe.side=left}{i,o}
   \fmflabel{$A,\mu$}{i}
   \fmflabel{$B,\nu$}{o}
\end{fmfchar*}}~~~~&~~~&~~~~~ \frac {i}{p^2} \delta_{AB}
\end{eqnarray}

\vskip .4cm
\noindent{\bf 3--gluon vertex}. The external gluons carry labels $(A,\mu,p)$, 
$(B,\nu,q)$ and $(C,\lambda,k)$ for the Lie algebra, momentum and 
Lorentz indices and are ordered in anticlockwise sense: 
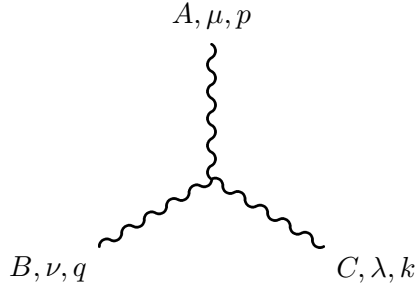
\vspace{5mm}

\begin{center}
\begin{fmfchar*}(85,85)
   \fmftop{i1}
   \fmfbottom{o1,o2}
   \fmf{photon}{i1,v1}
   \fmf{photon}{o1,v1}
   \fmf{photon}{o2,v1}
   \fmflabel{$A, \mu, p$}{i1}
   \fmflabel{$C, \lambda, k$}{o2}
   \fmflabel{$B, \nu, q$}{o1}
\end{fmfchar*}
\end{center}
\be
-g\left(f_{ABC}\,\cos (p\times q) + 
d_{ABC}\,\sin (p\times q)\right)\left(g_{\mu\nu}\,(p-q)_\lambda
+ g_{\nu\lambda}\,(q-k)_\mu+\,g_{\lambda\mu}(k-p)_\nu\right)\label{3gluv}
\ee

\vspace{5mm}
\noindent{\bf ghost vertex}. The gluon carries label $(A,\mu,k)$, the ghosts $(B,p)$
and $(C,q)$:
  
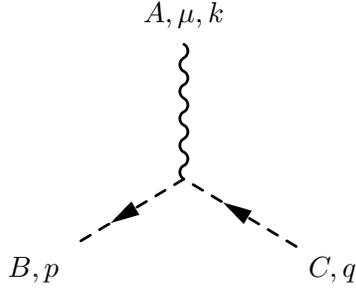
\begin{center}
\vspace{5mm}
\begin{fmfchar*}(85,85)
   \fmftop{i1}
   \fmfbottom{o1,o2}
   \fmf{photon}{i1,v1}
   \fmf{scalar}{v1,o1}
   \fmf{scalar}{o2,v1}
   \fmflabel{$A, \mu,k$}{i1}
   \fmflabel{$C, q$}{o2}
   \fmflabel{$B, p$}{o1}
\end{fmfchar*} 
\be
-g\,p_\mu \left(f_{ABC}\,\cos (p\times q) - d_{ABC}\,\sin (p\times q)\right)
\label{ghostv}
\ee
\end{center}

\vspace{5mm}
\noindent {\bf 4--gluon vertex}.  The gluons carry labels $(A,\mu,p)$,
$(B,\nu,q)$, $(C,\rho,r)$ and $(D,\sigma,s)$ for Lie algebra, Lorentz index 
and momentum. They are clockwise ordered:
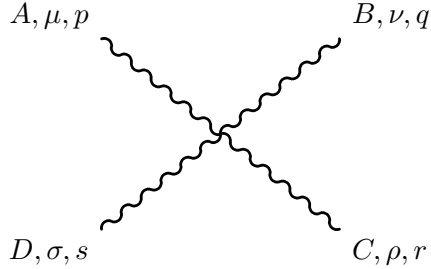
\vspace{6mm}

\begin{center}
\begin{fmfchar*}(90,90)
   \fmftop{i1,i2}
   \fmfbottom{o1,o2} 
   \fmf{photon}{i1,v1}
   \fmf{photon}{o1,v1}
   \fmf{photon}{o2,v1}
   \fmf{photon}{i2,v1}
   \fmflabel{$A, \mu, p$}{i1}
   \fmflabel{$D, \sigma, s$}{o1}
   \fmflabel{$B, \nu, q$}{i2}
   \fmflabel{$C, \rho, r$}{o2}
\end{fmfchar*}
\end{center}

\bea   
-ig^2\,&\Big[&\left(f_{ABX}\,\cos (p\times q) + 
d_{ABX}\,\sin (p\times q)\right)\no\\
&&\quad\quad\cdot\left(f_{XCD}\,\cos (r\times s) + d_{XCD}\,\sin (r\times s)\right)
(g_{\mu\rho}\,g_{\nu\sigma}- g_{\mu\sigma}\,g_{\nu\rho})\no\\
&+&\left(f_{ACX}\,\cos (p\times r) + d_{ACX}\,\sin (p\times r)\right)\no\\
&&\quad\quad\cdot\left(f_{XDB}\,\cos (s\times q) + d_{XDB}\,\sin (s\times q)\right)
(g_{\mu\sigma}\,g_{\nu\rho}- g_{\mu\nu}\,g_{\rho\sigma})\no\\
&+&\left(f_{ADX}\,\cos (p\times s) + d_{ADX}\,\sin (p\times s)\right)\no\\
&&\quad\quad\cdot\left(f_{XBC}\,\cos (q\times r) + d_{XBC}\,\sin (q\times r)\right)
(g_{\mu\nu}\,g_{\rho\sigma}- g_{\mu\rho}\,g_{\nu\sigma})\Big]\no
\eea
With elementary manipulations we can rewrite this as follows:
\bea
-\frac i4 g^2 &\Big[&
\Big( \cos(p\times s -q\times r)\, L_{ABCD} +  \sin(p\times s-
q\times r)\, M_{ABCD}\Big)\,T_{\mu\nu\rho\sigma} \no\\
&+&\Big( \cos(p\times r -q\times s) \,L_{BACD} -   
 \sin(p\times r -q\times s)\, M_{BACD}\Big)\,T_{\nu\mu\rho\sigma} \no\\
&+&\Big( \cos(p\times s +q\times r)\, L_{ACBD} + 
 \sin(p\times s+
q\times r)\, M_{ACBD}\Big)\,T_{\mu\rho\nu\sigma} \Big]
\label{4gluv}
\eea
The tensors $M,L,T$ are defined in the text.

The Feynman rules for $U(1)$ are formally obtained from the above ones by 
setting $\theta=0$, the tensor $f=0$ and $d=1$ (therefore, in particular,
$L=2$, $M=0$).

{\bf Acknowledgements} We would like to thank T.Krajewski, M.Sheikh--Jabbari
and A.Tomasiello for very long and useful discussions on the subject of this 
paper. We acknowledge as well discussions we had with R.Iengo, 
S.Terna and D.Zanon, and we thank A.Armoni for his comments on our manuscript. 
We thank in particular M.Schnabl for pointing out to us an error
in eq. (2.12) of the previous version of this paper.
 This work was partially supported by the Italian MURST for the 
program ``Fisica Teorica delle Interazioni Fondamentali''.

\end{fmffile}
\end{document}